\documentclass[11pt,twoside]{book}
\usepackage{vshalo}
\usepackage{longtable}
\usepackage{amsmath,amssymb}
\usepackage{graphicx}
\usepackage{lscape}
\usepackage{psfig}
\usepackage{epsf}
\usepackage{index}
\usepackage{natbib}
\usepackage{bigdelim}
\usepackage{multirow}
\makeindex
\newindex{aut}{adx}{and}{\Large Author  Index}
\newcommand{\axindex}[1]{\index[aut]{#1}}

\begin{document}

\pagestyle{myheadings}
\setcounter{equation}{0}\setcounter{figure}{0}\setcounter{footnote}{0}\setcounter{section}{0}\setcounter{table}{0}\setcounter{page}{149}
\markboth{Sterken, Samus \& Szabados}{VS-Halo Papers}
\title{A New Family of Models for Spherical Stellar Systems}
\author{Natalya Raspopova, Leonid Ossipkov}
\axindex{Raspopova, N. }  \axindex{Ossipkov, L.}
\affil{Saint Petersburg State University,  Saint Petersburg, Russia}

\begin{abstract}
A new two-parametric family of mass distribution for spherical stellar systems is considered. It generalizes families by \citet{KV72} and by \citet{AE06}. Steady velocity dispersions are found for these models by solving an equation of hydrostatic equilibrium. Axisymmetric generalizations of the model are discussed.
\end{abstract}

\section{Introduction}

In this study we suggest a new family of spherical mass distribution models that generalizes models by An \& Evans (2006, hereafter AE) and models by Kuzmin et al. \citep{KV72,KM69}. The family depends on two structural parameters. It includes Plummer's spheres \citep{pl11}, H\'{e}non's isochrones \citep{Henon59} and the model by Hernquist \citep{lars90} as special cases.
\medskip

\section{The potential--density pair}

Let us consider the dimensionless potential
\begin{equation}\label{phi}
    \Phi(r) = \frac{\alpha}{\alpha-1+w(r)}, \quad w(r) = (1+\kappa\,r^p)^{1/p}, \quad \kappa=O(\alpha^p),
\end{equation}
Here $\alpha > 1 $ and $p > 0$ are structural parameters. If $\alpha = 2$, $0 < p < 2$ we obtain a model by AE, for $\alpha > 0$, $p = 2$ we have a model by  Kuzmin et al. (1969, 1972).

\begin{figure}[!h]
 \centerline{\hbox{\psfig{figure=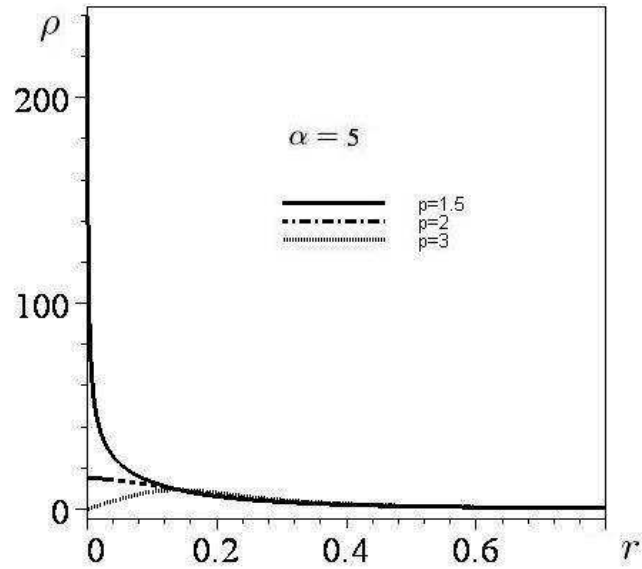,angle=0,clip=,width=9cm}}}
 \caption[]{The density profiles of the model for various values of $p$. }
\label{raspopova-fig1}
\end{figure}

\begin{figure}[!h]
 \centerline{\hbox{\psfig{figure=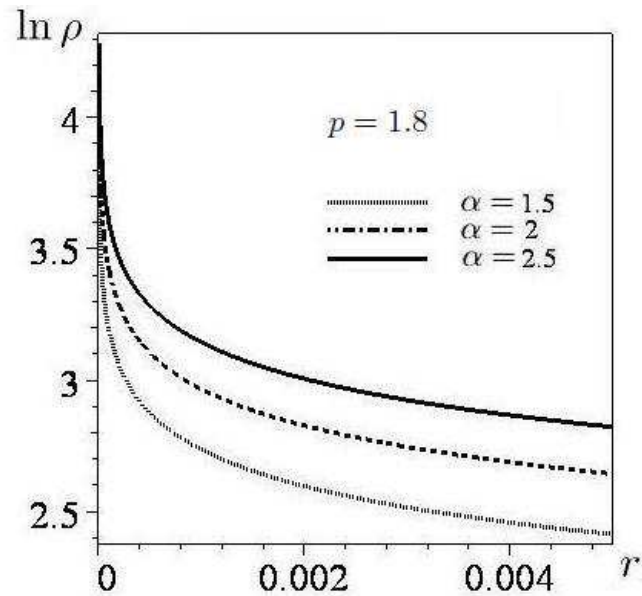,angle=0,clip=,width=9cm}}}
 \caption[]{The density profiles of the model for various values of $\alpha$. Note that all models are cusped, as $p<1$. }
 \label{raspopova-fig2}
\end{figure}

Poisson's equation yields the following expression for density
\begin{equation}\label{rho}
    \varrho(r) = \frac{\kappa}{\alpha}\,\frac{\Phi^2(r)\,r^{p-2}}{w(r)^{p-1}}\left[p + 1 - (p-1)\kappa\left(\frac{r}{w(r)}\right)^p - \frac{2\kappa}{\alpha}\frac{\Phi(r)\,r^p}{w(r)^{p - 1}}\right].
\end{equation}
It follows from (\ref{rho}) that our models with $0<p<2$ are cusped (as models by AE are). The density profiles for different values of parameters are shown in Figures \ref{raspopova-fig1}, \ref{raspopova-fig2}.

The circular speed is found to be
\begin{equation}\label{vc}
    v^2(r) = \frac{\kappa}{\alpha}\,\Phi^2(r)\frac{r^p}{w^{p - 1}(r)}.
\end{equation}


\section{Hydrostatics}

A run of velocity dispersion $\sigma^2_r$  can be found from an equation of hydrostatic equilibrium
\begin{equation}\label{hseq}
    \frac{1}{\varrho}\frac{d(\varrho\sigma^2_r)}{dr} + 2\beta\,\frac{\sigma^2_r}{r} = -\frac{GM(r)}{r^2}, \quad \beta = 1 - \frac{\sigma^2_t}{\sigma^2_r}.
\end{equation}

\begin{figure}[!h]
 \centerline{\hbox{\psfig{figure=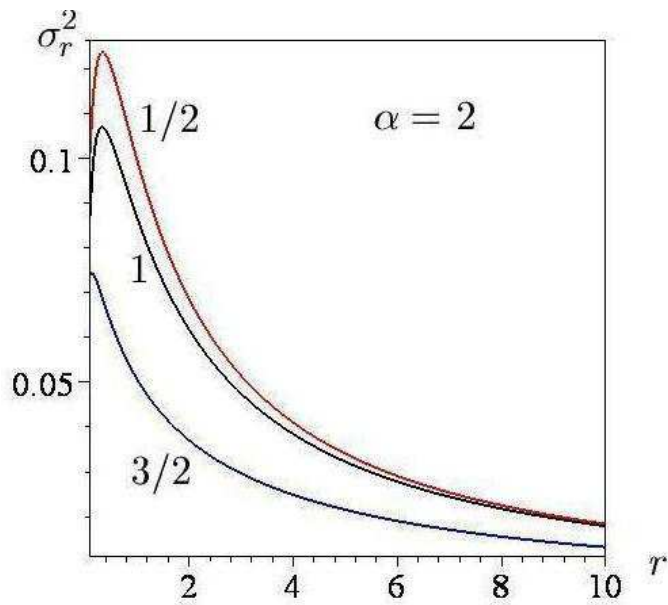,angle=0,clip=,width=9cm}}}
 \caption[]{Run of velocity dispersion for various values of $p$. }
 \label{raspopova-fig3}
\end{figure}

The results of calculations for an isotropic velocity distribution ($\beta=0$) are shown in Figure \ref{raspopova-fig3}. Central minima will appear in all models with density cusp. It can be obtained from (\ref{phi}) that
\begin{equation}\label{r_phi}
    r(\Phi)=\frac{\left[\left(\alpha-\left(\alpha-1\right)\Phi\right)^p-\Phi^p\right]^{1/p}}{\alpha\Phi}.
\end{equation}
Then it is possible to find an augmented density $\varrho(\Phi)$ and calculate an isotropic distribution function.

Stability of such models can be studied using the third Antonov law \citep{BTr}, namely, if $\displaystyle \frac{d^3\Phi}{d\varrho^3} > 0$ the model is stable against spherical perturbation. The validity of this inequality can be established after some laborious calculations.

\section{Axisymmetric generalizations}

Using the equipotential method by \citet{saklpo81} one can construct axisymmetric generalizations of the suggested model. We considered a potential of such models
\begin{equation}\label{gen}
    \varphi(R,z) = \varPhi(\xi), \quad \xi^2 = f(R,z),
\end{equation}
where $\varPhi(\xi)$ is the same function as \eqref{phi} and $f(R,z) = \mbox{const}$ is an equation of equipotential surfaces, $R$, $z$ being cylindrical coordinates. We considered the equipotentials by \citet{mmn75}:
\begin{equation}\label{mn}
\begin{array}{c}
    \xi^2 = R^2 + z^2 +  2\,(1-\varepsilon)\left(\sqrt{\varepsilon^2 + z^2} - \varepsilon\right),
\end{array}
\end{equation}
and by \citet{satoh80}:
\begin{equation}\label{satoh}
\begin{array}{c}
    \xi^2 = R^2 + z^2 +  \sqrt{4(1-\varepsilon)z^2+\varepsilon^2} - \varepsilon.
\end{array}
\end{equation}
Here $\varepsilon\in[0,1]$ is a new structure parameter. For spherical systems $\varepsilon=1$.

We found that for $\varepsilon$ close to $1$ the density is positive for $R$, $z$ everywhere. So we concluded that such non-spherical model can be used for approximating mass distribution in non-spherical star clusters and non-highly flattened galaxies.

\begin{figure}[!h]
 \centerline{\hbox{\psfig{figure=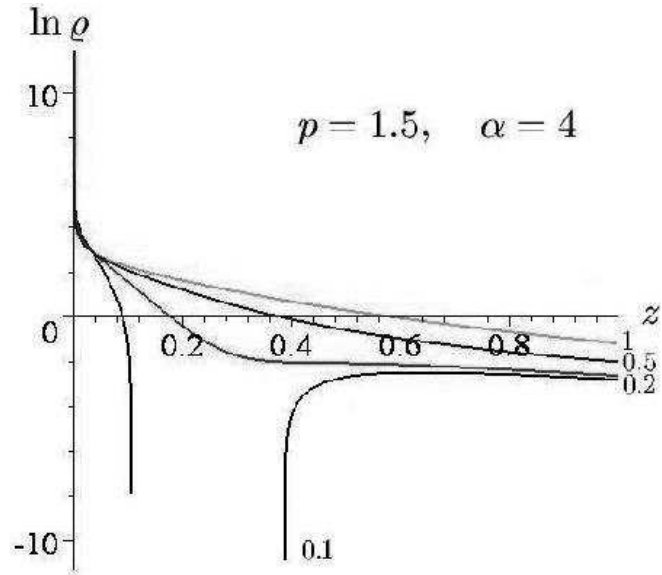,angle=0,clip=,width=9cm}}}
 \caption[]{The density profiles of the axisymmetric model with equipotentials by \citet {mmn75} for various values of $\varepsilon$. }
 \label{raspopova-fig4}
\end{figure}

\acknowledgments
 The work  was supported by RFBR under grant 08-02-00361.


\end{document}